\documentclass[]{spie}  
\usepackage{subcaption}

\usepackage{amsmath,amsfonts,amssymb}
\usepackage{graphicx}
\usepackage[colorlinks=true, allcolors=blue]{hyperref}

\title{DeepGI: An Automated Approach for Gastrointestinal Tract Segmentation in MRI Scans}

\author[*a]{Ye Zhang}
\author[b]{Yulu Gong}
\author[c]{Dongji Cui}
\author[d]{Xinrui Li}
\author[e]{Xinyu Shen}
\affil[a]{University of Pittsburgh, Pittsburgh, USA}
\affil[b]{Northern Arizona University, Flagstaff, USA}
\affil[c]{Trine University, Phoenix, USA}
\affil[d]{Cornell University, New York, USA}
\affil[e]{Columbia University, Frisco, USA}
\affil[*]{Corresponding author: yez12@pitt.edu}

\begin{document} 
\maketitle

\begin{abstract}
Gastrointestinal (GI) tract cancers pose a global health challenge, demanding precise radiotherapy planning for optimal treatment outcomes. This paper introduces a cutting-edge approach to automate the segmentation of GI tract regions in magnetic resonance imaging (MRI) scans. Leveraging advanced deep learning architectures, the proposed model integrates Inception-V4 for initial classification, UNet++ with a VGG19 encoder for 2.5D data, and Edge UNet for grayscale data segmentation. Meticulous data preprocessing, including innovative 2.5D processing, is employed to enhance adaptability, robustness, and accuracy.

This work addresses the manual and time-consuming segmentation process in current radiotherapy planning, presenting a unified model that captures intricate anatomical details. The integration of diverse architectures, each specializing in unique aspects of the segmentation task, signifies a novel and comprehensive solution. This model emerges as an efficient and accurate tool for clinicians, marking a significant advancement in the field of GI tract image segmentation for radiotherapy planning.
\end{abstract}

\keywords{Semantic Segmentation, Medical Image Segmentation, Inception-V4, UNet++, Edge UNet, 2.5D Data Processing}

\section{Introduction}
Gastrointestinal (GI) tract cancers pose a significant global health challenge, impacting millions annually. Radiotherapy, a crucial treatment modality, delivers targeted radiation to tumors while minimizing damage to surrounding healthy tissues. Advanced technologies like MR-Linacs (Magnetic Resonance Linear Accelerator systems) offer real-time imaging during treatment, allowing dynamic adjustments to radiation beams based on tumor and organ positioning.

Despite these advancements, the current standard practice in radiotherapy planning involves the manual delineation of the GI tract by radiation oncologists. This labor-intensive and time-consuming process is susceptible to inter-observer variability, hindering the efficiency of radiotherapy planning.

In response, this paper proposes an innovative solution employing deep learning techniques for the automated segmentation of the GI tract in magnetic resonance imaging (MRI) scans. The primary objective is to develop a model capable of accurately delineating the colon, small intestine, and stomach regions. The proposed model integrates sophisticated architectures, including Inception-V4 for initial classification, UNet++ with a VGG19 encoder for 2.5D data processing, and Edge UNet for grayscale data segmentation.

The methodology involves meticulous data preprocessing, incorporating 2.5D and grayscale processing to enhance the model's adaptability and robustness. The integrated segmentation architecture utilizes Inception-V4 for preliminary classification and UNet++ with VGG19 and Edge UNet for detailed segmentation. The strength of the model lies in its ability to provide automated, accurate, and efficient GI tract segmentation, addressing a critical gap in current radiotherapy planning practices.

The paper contributes to the field by:

\begin{enumerate}
    \item \textbf{Automating GI Tract Segmentation:} Introducing a model that automates the segmentation of the GI tract in MRI scans, significantly reducing the manual effort required in radiotherapy planning.

  \item \textbf{Integration of Advanced Architectures:} Leveraging state-of-the-art deep learning architectures, including Inception-V4, UNet++ with VGG19, and Edge UNet, to optimize accuracy in GI tract segmentation.

  \item \textbf{Innovative Data Preprocessing:} Implementing meticulous data preprocessing techniques, including 2.5D and grayscale processing, to enhance the model's adaptability and robustness to various imaging scenarios.

  \item \textbf{Efficiency Enhancement and Inter-Observer Variability:} Providing an efficient and accurate tool for clinicians to streamline the radiotherapy planning process, ultimately improving patient care and accessibility to advanced treatments while addressing inter-observer variability.
\end{enumerate}

The combination of these contributions positions the proposed model as a valuable advancement in the field of GI tract image segmentation, with potential implications for optimizing radiotherapy planning and improving patient outcomes.

\section{Related Work}
The dynamic field of medical image segmentation has witnessed a surge in research endeavors, particularly in the context of gastrointestinal cancers. Understanding the landscape of related work is paramount to contextualize the advancements made in our study. In the quest for automated segmentation of gastrointestinal organs, our exploration extends to a comprehensive review of existing methodologies and breakthroughs, serving as the bedrock upon which our research stands.

Kocak et al. \cite{kocak2023transparency} explores the application of deep learning in medical image segmentation, laying the foundation for subsequent advancements in the field.
Zhou et al.\cite{zhou2023artificial} Investigating the utilization of U-Net in medical imaging, this paper highlights its effectiveness in capturing intricate details while maintaining spatial context. Lu et al. \cite{lu2022decoupled} introduces a processor for analyzing power consumption data and enhancing modeling accuracy through comparisons between actual and converted power cycles. Edge U-Net incorporates innovations in edge detection using Holistically-Nested Edge Detection (HED), contributing to improved segmentation by capturing edge features.\cite{chen2023etu}.

Tianbo et al.\cite{tianbo2023bio} introduces a communication-free swarm intelligence system with robust functionality, integrating adaptive gain control, flocking SWARM algorithms, and object recognition for real-world applications. Zhang et al.\cite{zhang2023trep} Using a transformer module and deep evidential learning, TrEP outperforms existing models on pedestrian intent benchmarks.SuperCon presents a two-stage strategy for imbalanced skin lesion classification, emphasizing feature alignment and classifier fine-tuning, leading to state-of-the-art performance \cite{chen2022supercon}.

 Maccioni et al.\cite{maccioni2023magnetic}. provides insights into the challenges and advancements specific to segmenting gastrointestinal organs, considering anatomical variations and pathological conditions. Ronneberger et al.\cite{ronneberger2015u}  introduces the U-Net architecture, specifically designed for biomedical image segmentation, laying the foundation for subsequent advancements in the field.
The VGG architecture, with its deep convolutional networks, has significantly contributed to large-scale image recognition, forming the basis for subsequent image analysis models.\cite{simonyan2014very}.

Szegedy et al. \cite{szegedy2016rethinking} proposes improvements to the Inception architecture, enhancing accuracy while reducing computational costs, which has implications for various computer vision tasks. SegNet introduces an encoder-decoder architecture for image segmentation, contributing to the development of efficient segmentation models \cite{badrinarayanan2017segnet}.Y. Zhang et al. \cite{zhang2020manipulator}introduced a Machine Vision-based Manipulator Control System at the 2020 Cyber Intelligence Conference, enhancing robotic arms for precise object manipulation in two dimensions.

Zhang et al. \cite{zhang2023deep}(2023) introduce a novel deep learning model for breast cancer detection, achieving enhanced binary classification accuracy with a new pooling scheme and training method.
Liao et al.\cite{liao2020attention} (2020) propose the Attention Selective Network (ASN) for superior pose-invariant face recognition, achieving realistic frontal face synthesis and outperforming existing methods.J Lin's paper introduces a deep learning framework for Bruch's membrane segmentation in high-resolution OCT, aiding biomarker investigation in AMD and DR progression on both healthy and diseased eyes \cite{lin2022deep}.

T Xiao et al.\cite{xiao2022dual}. present dGLCN, a dual-graph learning convolutional network for interpretable Alzheimer's diagnosis,outperforming in binary classification on ADNI datasets with subject and feature graph learning.J Hu et al.\cite{hu2023m} present M-GCN, a multi-scale graph convolutional network for superior 3D point cloud classification on ModelNet40, emphasizing efficient local feature fusion.L Zeng et al. \cite{zeng2022graph}suggest a two-phase framework for Alzheimer's diagnosis, enhancing interpretability through weighted assignments in the graph convolutional network.S Chen et al.\cite{chen2022high} in Scientific Reports present a high-speed, long-range SS-OCT technology for anterior eye imaging with potential clinical applications.

S Chen et al.\cite{chen2023ultrahigh} in Ophthalmology use ultrahigh resolution OCT to differentiate early age-related macular degeneration from normal aging by detecting sub-RPE deposits.CL Quintana et al.\cite{quintana2022anterior} in Investigative Ophthalmology Visual Science propose an automatic method using ultrahigh-speed OCT for precise identification of external limbal transition points in scleral lens fitting.L Wang and W Xia in the Journal of Futures Markets propose an analytical framework for volatility derivatives, efficiently incorporating rough volatility and jumps \cite{wang2022power} .

While prior works have made valuable contributions, they often fall short in addressing the specific challenges posed by GI tract segmentation. Our study aims to fill these gaps by integrating advanced architectures, implementing innovative data preprocessing techniques, and leveraging the strengths of diverse models. This comprehensive approach offers a more tailored and effective solution for accurate GI tract image segmentation in medical imaging applications.

\section{Methodology}
\subsection{Overall Architecture}
\begin{figure}[h]
    \centering
    \includegraphics[width=0.6\textwidth]{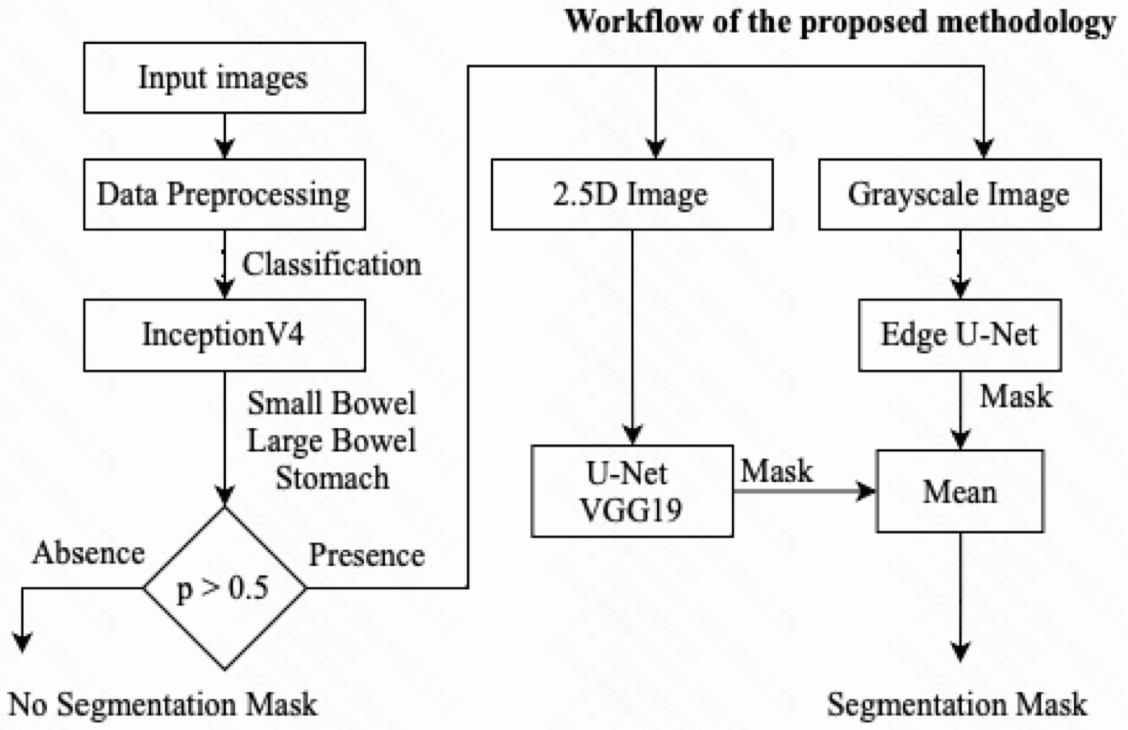}
    \caption{Overall Architecture}
    \label{fig:model_architecture}
\end{figure}

The proposed GI-Tract-Image-Segmentation model employs a tri-path approach, integrating advanced neural network architectures to achieve detailed segmentation. The overall architecture, depicted in Figure \ref{fig:model_architecture}, consists of three distinct pathways, each serving a specific purpose.

\subsubsection{Inception-V4 Pathway}
The first pathway initiates with Inception-V4, a state-of-the-art classification algorithm developed by the Google research team. Inception-V4 performs initial classification to identify healthy organs such as the colon, small intestine, and stomach in the input images. If no healthy organs are detected, the segmentation process concludes, generating a blank mask indicating the absence of segmentation.

\subsubsection{2.5D U-Net++ Pathway}
Simultaneously, the second pathway involves 2.5D data processing, incorporating depth information by stacking three consecutive MRI slices to create a 2.5D representation. The processed data is then input into the U-Net++ architecture, where the encoder utilizes VGG19. This pathway focuses on capturing detailed features in the segmented regions.

\subsubsection{Grayscale Edge U-Net Pathway}
The third pathway processes input images in grayscale and employs Edge U-Net for segmentation. Edge U-Net integrates Holistically-Nested Edge Detection (HED) for enhanced edge detection. Grayscale data processing is chosen for its simplicity and efficiency in capturing essential information for segmentation tasks.

The predictions from the 2.5D U-Net++ and grayscale Edge U-Net pathways are combined with the output from Inception-V4. The ensemble approach, achieved through averaging these predictions, ensures a comprehensive and accurate delineation of the GI tract regions in the input images. This integration leverages the strengths of 2.5D data processing, grayscale data processing, and initial classification with Inception-V4, enhancing the accuracy and robustness of the segmentation.

\subsection{Data Preprocessing}
\begin{figure}[h]
    \centering
    \includegraphics[width=0.7\textwidth]{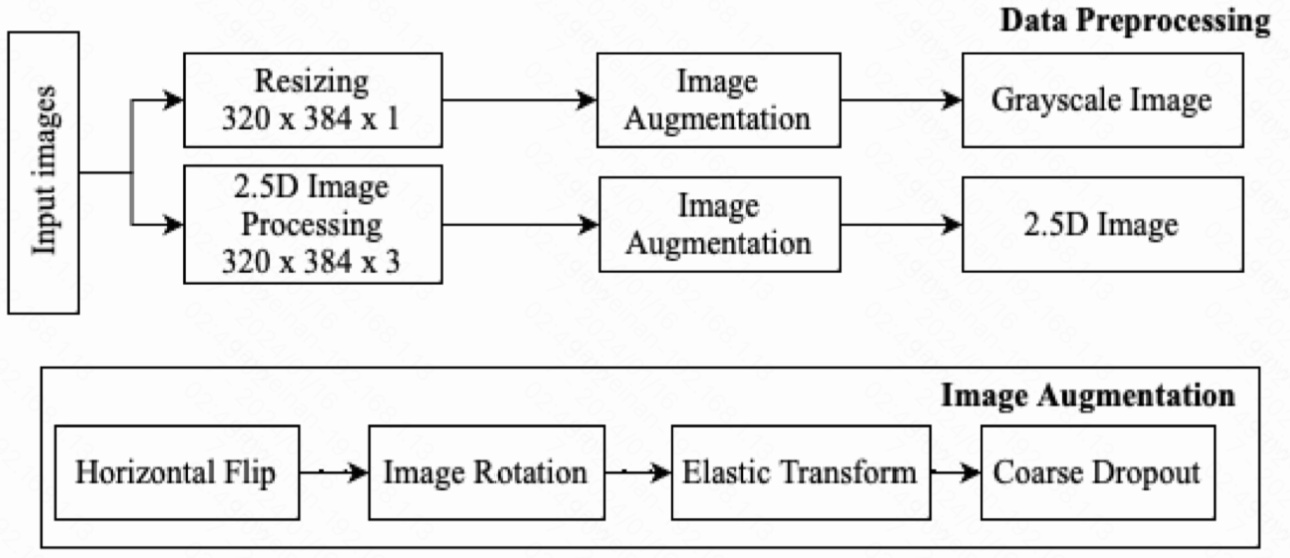}
    \caption{Data processing pipeline}
    \label{fig:data_processing_pipeline}
\end{figure}

Our data preprocessing pipeline, as shown in Figure \ref{fig:data_processing_pipeline} encompasses two distinctive processes tailored to enhance the model's adaptability and generalization.

\subsubsection{Spatial Augmentation Process}
The initial preprocessing step involves a spatial augmentation process designed to standardize and enhance the model's adaptability. Input images undergo resizing, ensuring uniformity at a resolution of 320x384 pixels through interpolation algorithms. Following resizing, a series of augmentation operations are applied to augment the dataset. These operations include horizontal flipping to simulate left-right mirror transformations, image rotation for improved directional robustness, elastic transformation mimicking non-linear distortions during image capture, and coarse random dropout simulating occlusions or missing data. The output of this process yields grayscale images, reducing color data dimensions for subsequent analysis and model training.

\subsubsection{Intensity Augmentation Process - Grayscale}
In addition to the spatial augmentation process, grayscale images undergo a dedicated intensity augmentation process. After resizing to a consistent resolution of 320x384 pixels, these grayscale images are subjected to a set of intensity adjustments. These adjustments aim to enhance the model's sensitivity to variations in pixel intensity, providing a nuanced understanding of grayscale features. The resulting grayscale images, enriched with enhanced intensity information, contribute to the overall diversity and robustness of the dataset for subsequent analysis and model training.

\subsubsection{2.5D Image Processing}
The second preprocessing process adopts 2.5D image processing techniques, enhancing traditional 2D image processing methods by introducing additional depth information. In this step, three consecutive MRI slices are stacked to generate 2.5D images simulating 3D volumetric data. This technique, validated in previous studies, preserves flat features while introducing spatial context between slices, providing richer contextual information for the model. Similar to the first preprocessing process, these images undergo the same augmentation operations to increase dataset diversity and improve model training effectiveness.

\begin{figure}[h]
    \centering
    \begin{subfigure}{0.3\textwidth}
        \includegraphics[width=\linewidth]{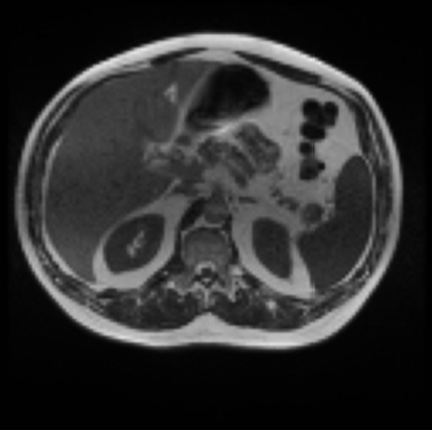}
        \caption{Original}
        \label{subfig:mask1}
    \end{subfigure}
    \hfill
    \begin{subfigure}{0.3\textwidth}
        \includegraphics[width=\linewidth]{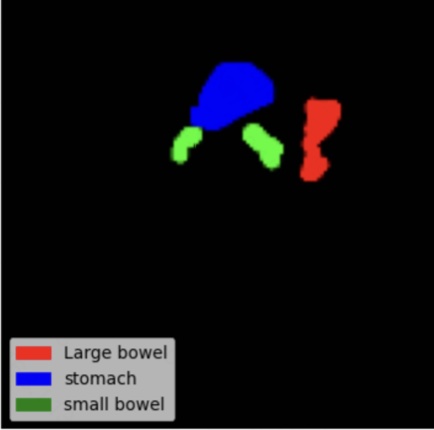}
        \caption{Masks}
        \label{subfig:mask2}
    \end{subfigure}
    \hfill
    \begin{subfigure}{0.3\textwidth}
        \includegraphics[width=\linewidth]{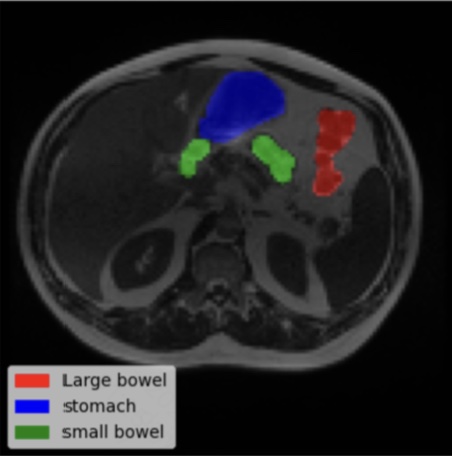}
        \caption{MRI with Masks}
        \label{subfig:mask3}
    \end{subfigure}
    \caption{Label Visualization}
    \label{fig:mri_overlay_combined}
\end{figure}

Due to 16-bit RLE-encoded masks in the training labels, visual analysis is challenging. Masks are converted to pixels, and marked regions are highlighted, as shown in Figure \ref{fig:mri_overlay_combined}.

\subsection{Model Architectures}

\subsubsection{Inception-V4 for Initial Classification}
Inception-V4, a product of Google's research team, assumes the role of the initial classifier in our model. The architecture prioritizes accuracy while mitigating computational costs through the utilization of smaller CNN sequences in lieu of more intricate structures. The initial block comprises filters of varying sizes (1x1, 3x3, and 5x5) and a max-pooling layer with a 2x2 filter size. Furthermore, the incorporation of batch normalization and residual connections contributes to additional performance optimization.

\subsubsection{U-Net++ with VGG19 Encoder}
The image segmentation model combines the U-Net architecture with VGG19 as the encoder, creating a synergistic U-Net++ framework. In this configuration, VGG19 functions as the encoder, extracting features from the input image, while U-Net++ orchestrates the image segmentation process. Within U-Net++, the encoder comprises convolutional layers from VGG19, and the decoder consists of multiple convolutional and upsampling layers, intricately connected to the encoder through skip connections. These skip connections play a crucial role in enabling the decoder to leverage the high-resolution features from the encoder for precise image segmentation.

\subsubsection{Edge U-Net}
\begin{figure}[h]
    \centering
    \includegraphics[width=0.8\textwidth]{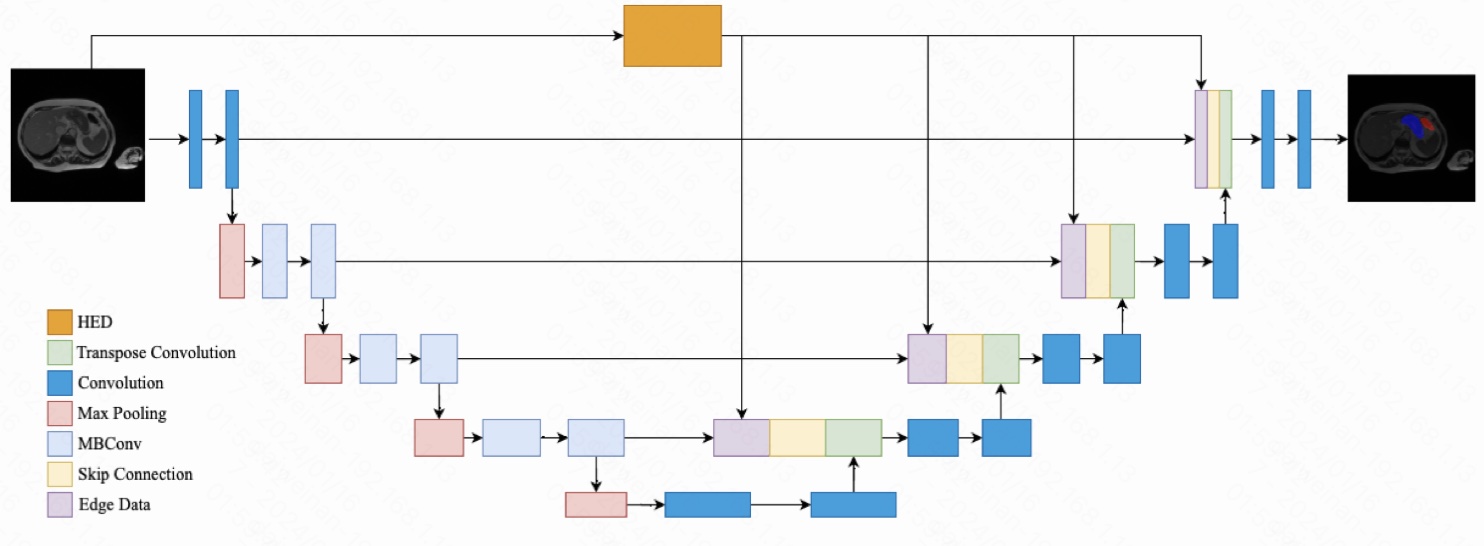}
    \caption{Edge U-Net Architecture}
    \label{fig:edge_unet_structure}
\end{figure}

Edge U-Net integrates edge-aware segmentation by substituting U-Net's encoder convolutional blocks with MBconv blocks, as shown in Figure \ref{fig:edge_unet_structure}. Additional skip connections are introduced to capture edge data from the input image. Edge detection is performed using the Holistically-Nested Edge Detection (HED) method, generating multi-scale edge images that are fused with the final edge features after upsampling.

\subsection{Evaluation Metrics}

\subsubsection{Dice Coefficient (DC)}

The Dice Coefficient, a fundamental metric in segmentation tasks, quantifies the similarity between predicted masks (PM) and ground truth masks (OM). It is defined as:

\begin{equation}
    DC(PM,OM) = \frac{2 \times |PM \cap OM|}{|PM| + |OM|}
\end{equation}

This metric provides a comprehensive assessment of the overlap between predicted and ground truth masks, offering insights into the segmentation accuracy.

\subsubsection{3D Hausdorff Distance}

To evaluate the spatial dissimilarity between two 3D masks, we employ the 3D Hausdorff Distance. This metric measures the maximum separation between corresponding pixels in the predicted mask (PM) and the ground truth mask (OM):

\begin{equation}
    HD(PM,OM) = \max \left( \max_{pm,om} \min(|pm - om|) \right)
\end{equation}

The 3D Hausdorff Distance serves as a crucial metric for capturing spatial discrepancies between predicted and ground truth masks, providing insights into segmentation robustness.

\subsubsection{Composite Score}

The final score is computed as a weighted combination of the Dice Coefficient and 3D Hausdorff Distance:

\begin{equation}
    \text{Score} = 0.4 \times \text{Dice Coefficient} + 0.6 \times \text{3D Hausdorff Distance}
\end{equation}

This composite score leverages both metrics, offering a balanced evaluation that considers both overlap accuracy and spatial dissimilarity. The combination provides a holistic assessment of the segmentation performance, capturing nuances that individual metrics might overlook.

\section{Experimental Results}

Our experiments aimed at evaluating different network models with diverse encoders for GI tract image segmentation on both grayscale and 2.5D preprocessed datasets. The findings and conclusions from these experiments are outlined below:

\subsection{Grayscale Image Segmentation}

In the context of grayscale images, a thorough evaluation of various network models employing distinct encoders was conducted. The validation scores are summarized in Table 
\ref{tab:grayscale_results}.
\begin{table}[h]
    \centering
    \begin{tabular}{|c|c|c|}
        \hline
        \textbf{Model} & \textbf{Encoder} & \textbf{Validation Score} \\
        \hline
        UNet & ResNet50 & 0.71599 \\
        UNet & Inception-V4 & 0.71002 \\
        UNet & Xception & 0.73761 \\
        UNet & EfficientNet-B0 & 0.68033 \\
        UNet & VGG19 & 0.78925 \\
        \hline
        UNet++ & ResNet50 & 0.7899 \\
        UNet++ & Inception-V4 & 0.80095 \\
        UNet++ & Xception & 0.79711 \\
        UNet++ & EfficientNet-B0 & 0.71372 \\
        UNet++ & VGG19 & 0.80717 \\
        \hline
        Edge UNet & - & 0.84046 \\
        \hline
    \end{tabular}
    \caption{Grayscale Image Segmentation Results}
    \label{tab:grayscale_results}
\end{table}

The results underscore that Edge UNet exhibits the most effective performance for grayscale image segmentation, outperforming other models in this domain.

\subsection{2.5D Image Segmentation}

In the realm of 2.5D images, the focus was on UNet++ with various encoders, with the following validation scores, as shown in Table \ref{tab:2.5d_results}.

\begin{table}[h]
    \centering
    \begin{tabular}{|c|c|c|}
        \hline
        \textbf{Model} & \textbf{Encoder} & \textbf{Validation Score} \\
        \hline
        UNet++ & ResNet50 & 0.80138 \\
        UNet++ & Xception & 0.7961 \\
        UNet++ & VGG19 & 0.84984 \\
        \hline
    \end{tabular}
    \caption{2.5D Image Segmentation Results}
    \label{tab:2.5d_results}
\end{table}

The validation scores affirm that UNet++ with VGG19 as the encoder excels in the domain of 2.5D images, showcasing superior segmentation performance.

\section{Conclusion}
This paper introduces a novel approach to automate the segmentation of gastrointestinal (GI) tract regions in MRI scans for radiotherapy planning. The proposed model, integrating advanced deep learning architectures such as Inception-V4, UNet++ with VGG19 encoder, and Edge UNet, addresses the manual and time-consuming segmentation process in current radiotherapy planning. The model's ability to provide automated, accurate, and efficient GI tract segmentation marks a significant advancement in the field, offering a valuable tool for clinicians to streamline the planning process and improve patient care.

The experimental results demonstrate the effectiveness of the proposed model, with Edge UNet performing exceptionally well in grayscale image segmentation, and UNet++ with VGG19 excelling in the domain of 2.5D images. These findings highlight the versatility and robustness of the model in handling different types of input data. Overall, this work contributes to the ongoing efforts in medical image segmentation, specifically addressing challenges related to GI tract segmentation, and provides a promising solution for enhancing radiotherapy planning efficiency.

\bibliography{report} 
\bibliographystyle{spiebib} 

\end{document}